\documentclass[fleqn,usenatbib]{mnras}

\usepackage[T1]{fontenc}

\DeclareRobustCommand{\VAN}[3]{#2}
\let\VANthebibliography\thebibliography
\def\thebibliography{\DeclareRobustCommand{\VAN}[3]{##3}\VANthebibliography}

\usepackage{graphicx}	
\usepackage{amsmath}	
\usepackage{amssymb}	
\usepackage{newtxtext,newtxmath}

\title[Hen 3-860: New eclipsing symbiotic binary]{Hen 3-860: New southern eclipsing symbiotic star observed in the outburst}

\author[J. Merc et al.]{
J. Merc,$^{1,2}$\thanks{E-mail: jaroslav.merc@student.upjs.sk}
R. G{\'a}lis,$^{2}$
M. Wolf,$^{1}$
P. Velez,$^{3}$
T. Bohlsen,$^{3}$
and B. N. Barlow$^{4}$
\\
$^{1}$Astronomical Institute, Faculty of Mathematics and Physics, Charles University, V Hole\v{s}ovi\v{c}k{\'a}ch 2, 180 00 Prague, Czech Republic\\
$^{2}$Institute of Physics, Faculty of Science, P. J. \v{S}af{\'a}rik University, Park Angelinum 9, 040 01 Ko\v{s}ice, Slovak Republic\\
$^{3}$Astronomical Ring for Amateur Spectroscopy Group\\
$^{4}$Department of Physics, High Point University, One University Parkway, High Point, NC 27268, USA\\
}

\date{Accepted 2021 November 28. Received 2021 November 22; in original form 2021 June 7}

\pubyear{2021}

\begin{document}
\label{firstpage}
\pagerange{\pageref{firstpage}--\pageref{lastpage}}
\maketitle

\begin{abstract}
Brightening of Hen 3-860, previously classified as an H$\alpha$ emitter, was detected by the ASAS-SN survey at the end of the year 2016. We have obtained the first spectroscopic observations of the transient and supplemented them with photometric data from the DASCH archive of astronomical plates, ASAS and ASAS-SN surveys. 
Based on the results of our analysis, we can classify the object as a~classical symbiotic star of the infrared type S, consisting of an M2-3 giant with $T_{\rm g}$\,$\sim$\,3\,550\,K, a radius \mbox{$R_{\rm g}\sim$\,60 - 75\,R$_{\sun}$,} and a luminosity $L_{\rm g}\sim$\,540 - 760\,L$_{\sun}$, and a hot and luminous component ($T_{\rm h}$\,$\sim$ $1-2 \times 10^5$\,K and $L_{\rm h}\,\sim 10^3\,\rm L_\odot$). The system experienced at least four outbursts in the last 120 years. In addition to the outbursts, its light curves revealed the presence of eclipses of the hot component and its surrounding (relatively cool) shell, which developed during the outburst and redistributed a fraction of the radiation of the hot component into the optical, by the giant, classifying the object as a representative of a group of eclipsing symbiotic stars. The eclipses allowed us to reveal the orbital period of the system to be 602\,days.
\end{abstract}




\section{Introduction}
\label{sec:intro}
Symbiotic stars are open binaries typically consisting of a cool giant of spectral type M (or K), a hot component and a complex circumbinary nebula. In most cases, the hot component is a white dwarf surrounded by extensive pseudophotosphere whose high quiescent temperature ($\sim$\,$10^{5}$\,K) and luminosity ($\sim$\,$10^{2-4}$\,L$_\odot$) is due to stable thermonuclear burning of hydrogen-rich matter accreted from the wind of the cool component \citep[see, e.g., reviews by][]{2012BaltA..21....5M,2019arXiv190901389M}. 


The symbiotic systems show a variety of photometric and spectroscopic changes on various timescales. The most prominent is the variability connected with the outbursts of these strongly interacting binaries. These significant changes in brightness usually attract the attention of observers, and it is often that previously unnoticed objects are then classified as symbiotic binaries. The thermonuclear outbursts of symbiotic novae typically have a large amplitude (several magnitudes), and can either last only for a few days and repeat themselves in the interval of years or decades (in the case of symbiotic recurrent novae), or the rise to the maximum takes years and the decline to quiescent magnitude could last for decades (in the case of 'slow' symbiotic novae), see, e.g., \citet[][]{2010arXiv1011.5657M,2013IAUS..281..162M}. 

In contrast, classical symbiotic binaries (Z~And-type) manifest series of individual brightenings by 1 - 3 mag (in $B$, with amplitude decreasing towards red) on the timescales of several months during the active stages. Their outbursts are far too frequent to be caused by nova-like thermonuclear runaways like those in symbiotic recurrent novae and seem to be caused either by the release of potential energy from extra-accreted matter or a shift of the hot component emission to longer wavelengths due to its expansion in response to increased mass accretion rate \citep[see the discussion in][]{2019arXiv190901389M}.


These interacting systems show also typical quiescent variability. Sinusoidal variations with the orbital period are usually observable at shorter wavelengths and are caused by apparent changes in the nebular radiation as a function of the orbital phase. According to the STB ionisation model \citep[][]{1984ApJ...284..202S}, the emission arises from the densest, partially optically thick parts of the cool giant wind photoionised by the hot component. At longer wavelengths, the light curves are typically dominated by the pulsations of the cool components \citep[Mira or semi-regular pulsations;][]{2003ASPC..303...41W, 2009AcA....59..169G, 2013AcA....63..405G}.

In some symbiotic systems, if the inclination is close to 90$^{\circ}$, the eclipses of the hot component (the white dwarf with its pseudophotosphere) and the surrounding, partially optically thick nebula by the giant can be detected in the light curves. These minima are especially prominent during the active stages and the transition period to the quiescence \citep[e.g.,][]{2001A&A...367..199S}. During these periods the pseudophotosphere of the white dwarf, in contrast with quiescence, expands and emits also in the optical part of the spectrum due to its lower temperature.



The star Hen 3-860 (= WRAY 15-10622; $\alpha_{\rm 2000}$ = 13:06:12.93, $\delta_{\rm 2000}$ = -53:22:52.50) was previously classified as an H$\alpha$ emitter by \citet{1966PhDT.........3W} and \citet{1976ApJS...30..491H}. The object was included in the International Variable Star Index database \citep[VSX;][]{2006SASS...25...47W} as a symbiotic candidate in November 2018 by an amateur astronomer Gabriel Murawski. This classification was based on the peculiar light curve of Hen 3-860 from the ASAS-SN survey \citep{2014ApJ...788...48S,2017PASP..129j4502K}, showing recent brightening (starting at the end of 2016) resembling a symbiotic outburst.

We have included the object in our observational campaign \citep[][]{2021MNRAS.506.4151M} focused on never spectroscopically observed and/or very poorly studied symbiotic candidates selected from the New Online Database of Symbiotic Variables \citep{2019RNAAS...3...28M}. For the selected symbiotic candidates, we have obtained optical spectroscopic data as well as collected the available photometry and other information from the literature in order to confirm or reject the symbiotic nature of the objects. Here, we report the results of the analysis of our low- and high-resolution spectra of Hen 3-860 and provide details on its long-term photometric behaviour.

The article is organised as follows: in Section \ref{sec:obs_data} we present the observational data used in the analysis. The symbiotic classification of the source is discussed in Section \ref{sec:symb}. The outburst activity and long-term photometry is described in Section \ref{sec:activity}, and parameters of the binary are presented in Section \ref{sec:bin}.

\section{Observational data}
\label{sec:obs_data}
The low-resolution spectroscopic observations of Hen 3-860 were obtained as a part of our campaign focused on symbiotic candidates carried out in cooperation with the ARAS Group\footnote{https://aras-database.github.io/database/symbiotics.html} \citep[\textit{Astronomical Ring for Amateur Spectroscopy};][]{2019CoSka..49..217T}. The first low-resolution spectrum was obtained at JD 2\,459\,049.4 (July 18, 2020) at the iTelescope facility at Siding Springs, Australia using a~Planewave~CDK 12.5" telescope equipped with a LISA spectroscope and an Atik 460EX camera. The second spectrum was secured at JD 2\,459\,051.5 (July 21, 2020) at a private station in Mirranook Armidale, Australia using an ASI183MM camera and a LISA spectroscope mounted on a~Celestron 11" telescope. A total of eight and four exposures of 600~s each, respectively, were obtained and summed on both occasions. The spectra have resolution R\,$\sim$1\,200 and R\,$\sim$1\,800, respectively.

Two high-resolution spectra (R $\sim$ 25\,000) were obtained with \mbox{CHIRON} cross-dispersed echelle spectrometer \citep{2013PASP..125.1336T} at the SMARTS 1.5-meter telescope located at the Cerro Tololo Inter-American Observatory, Chile. The two spectra were secured at JD 2\,459\,249.9 (February 4, 2021) and JD 2\,459\,294.6 (March 21, 2021) with exposure time of 1\,800\,s each. Basic information about the spectra used in this study (JD, date, resolution, spectral range, observer) are summarised in the log of observations (Table \ref{table:log_obs}).

The spectroscopic observations of Hen 3-860 were supplemented by available photometry obtained from the All-Sky Automated Survey \citep[ASAS; $V$ filter;][]{1997AcA....47..467P} covering JD 2\,452\,032 - 2\,454\,836 (May 02, 2001 - January 4, 2009), and All-Sky Automated Survey for Supernovae \citep[ASAS-SN; \textit{V}~and \textit{g} filters; ][]{2014ApJ...788...48S, 2017PASP..129j4502K} covering JD 2\,457\,423 - 2\,459\,331 (February 04, 2016 - April 26, 2021). The historical data, irregularly covering the interval of JD 2\,415\,115 - 2\,434\,127 (April 5, 1900 - April 24, 1952), were obtained from the DASCH (Digital Access to a Sky Century at Harvard) archive of digitised glass photographic plates of the Harvard College Observatory \citep{2010AJ....140.1062L}. Additional measurements in \textit{B, V, R$_c$}, and \textit{I$_c$} filters were obtained at JD 2\,459\,333.6 (April 29, 2021) at the Danish 1.54-meter telescope at La Silla, Chile.

To construct the multi-frequency SED of Hen 3-860, we have collected the data from \textit{Gaia} EDR3 \citep{2020arXiv201201533G}, SkyMapper \citep{2018PASA...35...10W}, APASS \citep{2015AAS...22533616H}, 2MASS \citep{2006AJ....131.1163S}, and WISE \citep{2010AJ....140.1868W}.

The study of the reddening and distance of the object are described in detail in Sections \ref{sec:redding} and \ref{sec:cool}.

\section{Symbiotic classification} 
\label{sec:symb}

\begin{figure}
\begin{center}
\includegraphics[width=\linewidth]{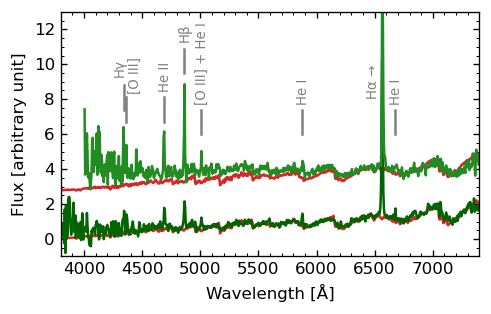}
\caption{Low-resolution spectra of Hen 3-860. The spectra obtained on July 18, 2020 and July 21, 2020 are depicted by the dark and light green colour, respectively. The best-fitting empirical spectrum (M3 III) from the MILES library \citep{2011A&A...532A..95F} is shown in red (see the text for details). The identification of prominent emission lines detected in the spectra is given by the vertical lines.}
\label{fig:spectra}
\end{center}
\end{figure}

\begin{figure*}
\begin{center}
\includegraphics[width=\textwidth]{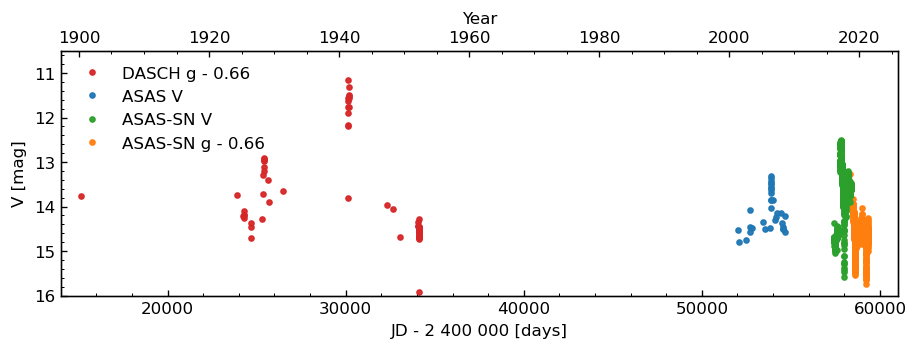}
\caption{Long-term light curve of Hen 3-860 showing 121 years (1900 - 2021) of the photometric history of this symbiotic system. The light curve was constructed on the basis of photographic observations from the DASCH archive and data from the ASAS and ASAS-SN surveys.}
\label{fig:long_term_LC}
\end{center}
\end{figure*}

\begin{figure}
\begin{center}
\includegraphics[width=\linewidth]{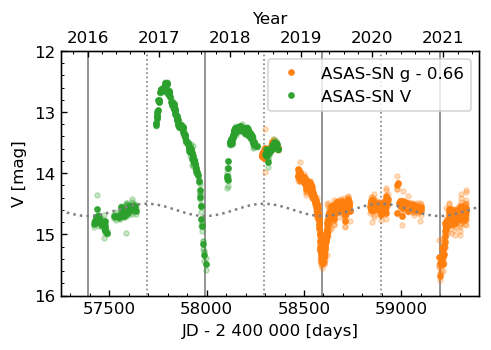}
\caption{Recent light curve of Hen 3-860 covering the interval of years 2015 - 2021. The light curve was constructed on the basis of ASAS-SN $V$ and $g$ observations. The grey dotted sinusoidal curve fits the brightness variation of the symbiotic star during its transient and quiescent periods (outside eclipses). Vertical grey solid and dotted lines show the position of the primary and secondary eclipses, respectively. The eclipse times were calculated using the ephemeris given by eq. \ref{eq:eph}.}
\label{fig:recent_LC}
\end{center}
\end{figure}

As was mentioned in the Introduction, the possible symbiotic classification of Hen 3-860 was proposed based on the peculiar light curve of the object from the ASAS-SN survey. To definitively confirm the symbiotic nature of an object, its quiescent optical spectrum has to satisfy the following criteria\footnote{Accreting-only symbiotic stars, which do not satisfy these criteria, are not discussed here, see, e.g., \citet{2021MNRAS.505.6121M} and \citet[][]{2021PhDT........17L} and references therein.} \citep{1986syst.book.....K,2000A&AS..146..407B}: (a) the presence of late-type giant absorption features, (b) the presence of the strong \ion{H}{i} and \ion{He}{i} emission lines, (c) the presence of emission lines with an ionisation potential of at least 35\,eV and an equivalent width exceeding 1\,\AA. The presence of the Raman-scattered \ion{O}{vi} lines \citep{1989A&A...211L..31S} is a~sufficient condition, even if the presence of the cool giant is not evident in the spectra. However, we should note that these lines are observed only in about half of the confirmed symbiotic systems \citep[see, e.g., the recent census by][]{2019ApJS..240...21A}. 

Our low-resolution optical spectra of Hen 3-860 (Fig. \ref{fig:spectra}) satisfy all these conditions. Namely, the spectra show the M2-3 III continuum (see also Sec. \ref{sec:cool}), and the emission lines of \ion{H}{i}, \ion{He}{i}, [\ion{O}{iii}], and \ion{He}{ii}. In addition, the high-resolution spectra confirmed the presence of [\ion{Fe}{vii}] lines at 5721\,\AA\,\,and 6087\,\AA\,\,with ionisation potential of 99\,eV. The symbiotic classification of Hen 3-860 is further supported by its position in the diagnostic diagram employing the [\ion{O}{iii}] and Balmer lines fluxes \citep[for more details, see ][and Fig. 1 therein]{2017A&A...606A.110I}. In the diagram, Hen 3-860 is located in the region occupied solely by symbiotic stars ([\ion{O}{iii}] $\lambda$5006\,/\,H$\beta$ = 0.27, [\ion{O}{iii}] $\lambda$4363\,/\,H$\gamma$ = 0.24).

The IR colours of Hen 3-860 also confirm its symbiotic classification, as they satisfy the IR criteria for symbiotic binaries presented by \citet{2019MNRAS.483.5077A,2021MNRAS.502.2513A}. The NIR colours of Hen 3-860 are typical for S-type (stellar) symbiotic binaries. Note that the majority of symbiotic binaries belong to this group \citep[][]{2000A&AS..146..407B,2019ApJS..240...21A}.

\section{Activity and long-term photometry} \label{sec:activity}

The historical light curve constructed on the basis of photographic observations from the DASCH archive of astronomical plates and data from the ASAS and ASAS-SN surveys is shown in Fig. \ref{fig:long_term_LC}. The ASAS-SN $g$ data were simply linearly shifted to correspond to the $V$ magnitudes. The DASCH data are calibrated using ATLAS-REFCAT2 \citep{2018ApJ...867..105T} and correspond to the $g$ magnitudes. For this reason, we have applied to this dataset the same shift as for the ASAS-SN $g$ data. Due to possible inaccuracies of these shifts, it is necessary to treat the particular historical magnitudes with caution. On the other hand, they allow us to compare the amplitudes of individual outbursts and analyse the long-term behaviour of Hen~3-860.

There are four outbursts apparent in the light curve that occurred in 1928, 1941, 2006 and 2016-2019. For the three outbursts detected in the past, only the lower limit of duration ($\sim$100, 100, and 150 days, respectively) could be estimated due to sparse time coverage. The recent outburst (shown in Fig. \ref{fig:recent_LC}), which started in 2016 and continued until 2019, lasted at least for $\sim$1\,000 days. Due to the seasonal gap in observations, it is not possible to determine the exact time when the outburst started. The system brightened by $\sim$1.6, 3.4, 1.2, and 2.1\,mag in $g$/$V$ during these outbursts. Interestingly, the first pair of the outbursts and the recent one repeated at about same interval of 11 - 13 years. The amplitude, duration and recurrence time of the outbursts is similar to the behaviour of classical symbiotic binaries \citep[e.g., AG~Dra, Z~And; ][]{2019CoSka..49...19S, 2019arXiv190901389M}, and therefore we can classify the object as a symbiotic binary of Z~And-type.

The quiescent orbital variability of symbiotic stars is typically well observable only at shorter wavelengths \citep[\textit{U}, \textit{B} filters;][]{1998A&A...335..545F,2009NewA...14..336S,2012JAVSO..40..572M} and its amplitude declines significantly toward longer wavelengths. On the other hand, pulsations of cool components in symbiotic binaries \citep[on the time scale of 50 -- 200 days in S-type symbiotics;][]{2013AcA....63..405G} are usually detectable only at longer wavelengths (\textit{R}, \textit{I}). 

\begin{figure}
\begin{center}
\includegraphics[width=0.945\linewidth]{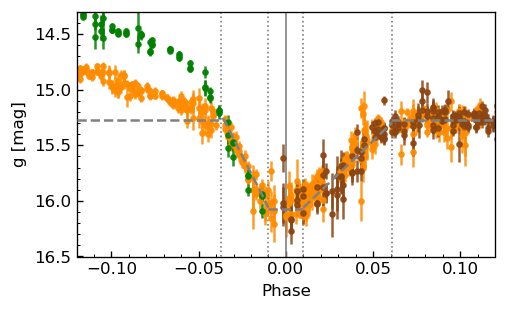}
\includegraphics[width=0.95\linewidth]{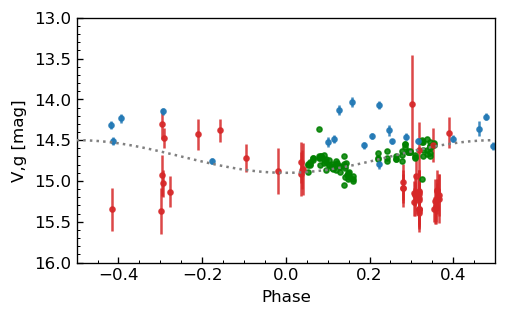}
\caption{Details of the light curves of Hen 3-860. Upper panel: Part of the phased light curve of Hen 3-860 showing the observed eclipses. The green, orange and brown points depict the eclipses observed in August 2017 (ASAS-SN $V$ shifted by +0.66 mag), April 2019 (ASAS-SN $g$) and December 2020 (ASAS-SN $g$ shifted by -0.14 mag), respectively. The dotted lines denote the contact times of the eclipse and the solid line mark the light minimum. The dashed line shows the simplified eclipse model. Out-of-eclipse differences are caused by the outburst. Lower panel: The phased light curve of \mbox{Hen~3-860} constructed using the observations obtained during the supposed quiescent periods of this symbiotic star. The phase diagram was constructed on the basis of photographic observations from the DASCH archive (red) and data from the ASAS (blue) and ASAS-SN (green) surveys. The dotted lines shows sinusoidal variability with the orbital period calculated using the ephemeris given by eq. \ref{eq:eph}.}
\label{fig:eclipses}
\end{center}
\end{figure}

Both, historical data from DASCH, and ASAS survey and recent ASAS-SN data of the object were obtained only in two filters with a~similar effective wavelength (\textit{g} has maximum efficiency between the Johnson \textit{B} and \textit{V} filters). Moreover, the coverage of historical data is rather sparse and the scatter is too large for reasonable analysis of quiescence variability of \mbox{Hen 3-860} (lower panel of Fig. \ref{fig:eclipses}). 

Although after the outburst covered by the ASAS-SN survey the optical brightness of Hen 3-860 faded nearly to the level of quiescent phase of this symbiotic binary, the presence of eclipses in the light curve suggests that the system is in the transition period from the recent activity to the quiescence. Therefore, there are no usable quiescent data for Hen 3-860 available. 

Nevertheless, in addition to the eclipses (see Sec. \ref{sec:orb} and upper panel of Fig. \ref{fig:eclipses}), the recent ASAS-SN light curve of Hen 3-860 revealed the presence of small changes in brightness. To visualise these variations, we fitted the light curve of the symbiotic system during this transition period (outside eclipses) by sinusoidal curve (see Fig. \ref{fig:recent_LC}). The mean magnitude and amplitude of these sinusoidal variation in the $g$ filter is around 15.3 mag and 0.1 mag, respectively. 

Moreover, our observations secured in April 2021 (\textit{B} = 15.85\,mag, \mbox{\textit{V} = 14.53\,mag}) differ significantly from that from the APASS catalogue (\textit{B} = 16.83\,mag, \textit{V} = 14.77\,mag) suggesting the possibility of quiescent variability with rather large amplitudes, at least at short optical wavelengths. This is further supported by the analysis of \citet{2021A&A...648A..44M} who included Hen 3-860 in their catalogue of large-amplitude variables based on the uncertainties of the \textit{Gaia} DR2 magnitudes. They obtained amplitude of 0.15 and 0.05\,mag in the \textit{Gaia} $BP$ and $RP$ bands for Hen 3-860, respectively. These amplitudes are consistent with the amplitude we obtained for the light curve in the $g$~filter taking into account that the \textit{Gaia} $BP$ filter has a lower, and $RP$ filter has a higher effective wavelength than the $g$~filter, and the amplitude of quiescent variability of symbiotic stars rises toward blue.

\section{Binary parameters} \label{sec:bin}

\subsection{Orbital period and the eclipses}\label{sec:orb}

In the recent light curve of Hen 3-860 (Fig. \ref{fig:recent_LC}), three primary eclipses could be detected. To evaluate the orbital period of the system, we have used the two eclipses which occurred after the recent outburst in April 2019 and December 2020. The eclipse observed during the outburst was not taken into account because there is a seasonal observational gap in the middle of the eclipse. In addition, during the outbursts of symbiotic binaries, the hot components change their size (see also the discussion below), which could distort the minimum time estimate causing the apparent changes in the O-C diagram \citep{1998A&A...338..599S}.

Using the two eclipses that occurred during the transition period from active to quiescent stage in April 2019 and December 2020, we have obtained the linear ephemeris for the photometric minimum of Hen 3-860:
\begin{equation}\label{eq:eph}
JD{\rm_{min}} = (2\,458\,594.8 \pm 3.2) + (602.1 \pm 4.8) \times E
\end{equation}

The derived value of the orbital period is well consistent with other S-type symbiotic stars \citep[e.g.,][]{2013AcA....63..405G}. We also indicated the times of possible secondary minima in Fig. \ref{fig:recent_LC} (dotted lines), assuming a circular orbit. At least in 2018 (during the outburst), a~decrease of $\sim$ 0.2 mag is noticeable at the time of the secondary minimum. 

The part of the phased light curve showing the primary eclipses of Hen 3-860 is depicted in the upper panel of Fig. \ref{fig:eclipses}. It is clearly seen that the shape of the eclipses is slightly asymmetric, the ingress to the eclipse taking shorter time (t$_2$ - t$_1$ = 16.3\,d) than ascent from it (t$_4$ - t$_3$ = 30.7\,d), pointing to the asymmetry of the eclipsed object (the relatively cool shell surrounding the hot component, which was created during the outburst and redistributed a fraction of the radiation of the hot component into the optical). Under the assumption of circular orbit and total mass of the system $\sim$\,2 - 3\,M$_\odot$ \citep[i.e., values typical for other symbiotic stars; see, e.g.,][]{2003ASPC..303....9M}, the binary separation corresponding to the obtained orbital period would be 1.8 - 2.0\,AU. Using the contact times, we can then estimate the lower limits of the sizes of cool component and the eclipsed object, assuming $i$ = 90$^{\circ}$: $R_{\rm g}$ = 57 - 63 $R_\odot$, R$_{\rm e}$ = 33 - 36 R$_\odot$, respectively. The derived radius of the giant is well consistent with a M2 III star \citep{1999AJ....117..521V}, consistent with the spectral type estimated from our low-resolution spectra (see Sec. \ref{sec:cool}). The size of the eclipsed object is similar to that of the known eclipsing symbiotic binary AX Per in transition period (Sec. \ref{sec:axper}).

The gap in the observations of the eclipse during the outburst precludes accurate estimates of contact times, but the eclipse duration suggests a several-fold increase in the size of the eclipsed object. This is consistent with the assumption that the hot component of symbiotic stars expands during outburst \citep[e.g.,][]{1986syst.book.....K,2011A&A...536A..27S}. We should note that the brightness of Hen 3-860 during the eclipse is close to the limit of the ASAS-SN survey which introduces inaccuracies into the determination of the contact times. Therefore, only precise (ideally multi-colour) long-term photometric observations will help to refine the findings presented in this section. 

\subsection{Reddening}
\label{sec:redding}
The total Galactic extinction in the direction of Hen 3-860, given by the map of \citet{2011ApJ...737..103S} is $E(B-V) = 0.45$. The values $E(B-V) = 0.4 - 0.5$ based on the interstellar \ion{Na}{i} \citep{2005A&A...434.1107M} are in agreement with the value of extinction given by the dust maps. Symbiotic stars (at least those of the D-type) are subjected to additional extinction due to circumstellar matter and the value given by the dust maps is often an underestimation of the true reddening. We have employed other methods widely used in the case of symbiotic stars to obtain the independent reddening estimates.

Using the comparison of intrinsic \citep{1988PASP..100.1134B} and observed $(J-K)$ colour of Hen 3-860 from the 2MASS catalogue \citep{2006AJ....131.1163S}, we have obtained $E(B-V) = 0.71$ and 0.62 for M2 and M3 giants, respectively (see Sec. \ref{sec:cool}). The values from the 2MASS catalogue were converted to the standard system of \citet{1988PASP..100.1134B} using the transformations of \citet{2001AJ....121.2851C}. Note that in the case of the S-type symbiotic star Gaia18aen, the extinction value based on the comparison of synthetic and observed NIR spectra of the cool giant was larger than the one from the dust maps \citep[by about 0.2 mag, similar to Hen 3-860;][]{2020A&A...644A..49M}.

We have also estimated the reddening from the emission-lines ratios following \citet[][]{1997A&A...327..191M}. In S-type symbiotics, the conditions differ from the case B recombination and the reddening-free ratio of \mbox{H$\alpha$\,/\,H$\beta$ $\sim$ 5 - 10} \citep[][]{1997A&A...327..191M}. Using these boundary values, we obtained $E(B-V) = 0.61 - 1.31$. Another useful ratio is that of \ion{He}{i} 7065\AA\,/\,\ion{He}{i} 5876\AA\,\,which approximately equals 0.84 in S-type symbiotic stars \citep{1994MNRAS.268..213P}. This value corresponds to the reddening $E(B-V) = 0.62$ in the case of \mbox{Hen 3-860}. Using the derived ratio of [Fe VII] emission lines ([\ion{Fe}{vii}] 5721\AA\,/\,[\ion{Fe}{vii}] 6087\AA = 0.65), which are visible only in our high-resolution spectra, we obtained $E(B-V) = 0.45$. This value is lower than the other estimates but this could be caused by a low S/N ratio of the continuum in the spectra used and the consequent uncertainty in the flux measurements.

All the reddening estimates assume the total-to-selective absorption ratio R$_{V}$ = 3.1 and the reddening law of \citet{1989ApJ...345..245C}. In this research, we have adopted the value $E(B-V) = 0.65$.

\subsection{Cool giant and the distance}
\label{sec:cool}

The presence of the cool giant in Hen 3-860 is revealed by the TiO bands well detectable in our low-resolution spectra. Using the TiO indices presented by \citet[][eqs. 1 and 2 therein]{1987AJ.....93..938K}, we have estimated the spectral type of the cool giant to be M2. We have also compared the observed spectra (Fig. \ref{fig:spectra}) to the ones from the MILES empirical library of stellar spectra \citep[][]{2011A&A...532A..95F} in order to confirm this spectral classification. For calculation of $\chi^2$, only the red part of the spectra (> 5\,500\,\AA), in which the giant dominates, was used. The empirical spectra were down-sampled to the resolution of the observed ones before the analysis. The best fit was obtained for a M3 giant. Using the statistical relations of \citet{2020RAA....20..139M}, we can estimate the effective temperatures corresponding to the obtained spectral types, namely 3\,590\,K and 3\,480\,K for M2 III and M3 III, respectively. 

\begin{figure}
\begin{center}
\includegraphics[width=\linewidth]{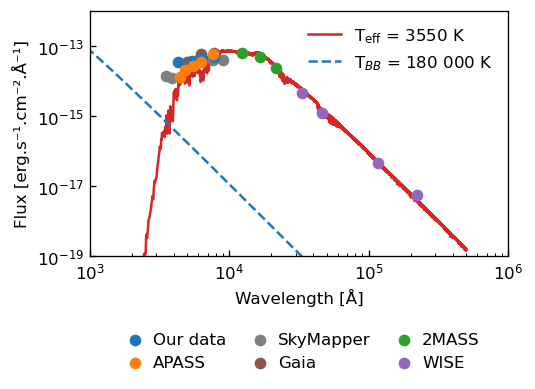}
\caption{Multi-frequency SED of Hen 3-860 constructed on the basis of our data obtained with the Danish 1.54-m telescope at La Silla, Chile and data from the \textit{Gaia} EDR3, SkyMapper, APASS, 2MASS, and WISE. The best-fitting theoretical spectrum with $T_{\rm eff}$ = 3\,550\,K and $\log g = 1.0$ is shown in red \citep{2014IAUS..299..271A}. The dashed blue line denotes the radiation of a black-body in a distance of 5\,kpc with a temperature of $1.8 \times 10^{5}\,$K, and luminosity of 1\,170\,L$_{\sun}$ (a hot component of Hen 3-860; see the text for details). We should note that the SED represents a mean spectrum of Hen 3-860 as the individual points were obtained over a relatively long period of time covering various phases of activity of this symbiotic star.}
\label{fig:sed}
\end{center}
\end{figure}

Similar temperature was obtained from the multi-frequency SED of Hen 3-860 (Fig. \ref{fig:sed}), which was compared with the BT-Settl grid of theoretical spectra \citep{2014IAUS..299..271A} downloaded from the Theoretical spectra webserver at the SVO Theoretical Model Services\footnote{http://svo2.cab.inta-csic.es/theory/newov2/index.php}. In this case, we have used only the APASS \textit{r'} and \textit{i'}, 2MASS and WISE observations, as at shorter wavelengths the contribution of the nebular radiation and the hot component is not negligible, and would artificially increase the estimate of the temperature of the cool component. Moreover, their contribution is changing during various phases of activity Hen 3-860 covered by observations used for construction of the SED. The best fit was obtained for the temperature of 3\,550\,K and $\log g = 1.0$ (corresponding to $\sim$ M2.5 giant), consistent with the estimates from the optical spectra.


The giant radius corresponding to the obtained spectral types is in the range of 60 - 75 R$_\odot$ \citep{1999AJ....117..521V}, which is consistent with the lower limits from the contact times of the eclipses of Hen 3-860 assuming $i$ = 90$^{\circ}$ (57 - 63 R$_\odot$; Sec. \ref{sec:orb}). The luminosity  can then be estimated directly from the Stefan-Boltzmann law. For the range of temperatures of 3\,500 - 3\,600\,K and radii 60 - 75\,R$_\odot$, one can obtain the luminosity of the giant $L_{\rm g}$ = 540 - 760\,L$_\odot$. We can use these values to estimate the distance to Hen 3-860. The range of absolute bolometric magnitudes of Hen 3-860 corresponding to the obtained values of the luminosity is $M_{\rm bol}$ = -(2.46 - 2.09)\,mag. Using the relations of bolometric correction $BC_{\rm K}$ and temperature presented by \citet{2010MNRAS.403.1592B}, we calculated $M_{\rm K}$ = -(5.39 - 4.93)\,mag. The dereddened \textit{K} magnitude of Hen 3-860 from the 2MASS catalogue is 8.15\,mag. From apparent and absolute \textit{K} magnitude, we calculate a distance of 4.1 - 5.1\,kpc. 


The ratio of the observed and intrinsic flux values can be used to determine the angular radius of the giant. For the measured bolometric flux of the giant corrected for interstellar extinction $F_{\rm g}^{\rm obs} = 9.3 \times 10^{-10}\,\rm erg\,s^{-1}\,cm^{-2}$ (red line in Fig. \ref{fig:sed}) and intrinsic bolometric flux of $7.2 \times 10^{9}\,\rm erg\,s^{-1}\,cm^{-2}$ obtained from the BT-Settl theoretical spectrum \citep{2014IAUS..299..271A} with $T_{\rm eff}= 3\,550$\,K and $\log g = 1.0$ we get the angular radius of the giant $\theta = 3.6 \times 10^{-10} = 7.4 \times 10^{-5}$\,arcsec. Under the assumption that the giant radius is according its derived spectral type in the interval of 60 - 75 R$_\odot$ we get its distance in the interval 3.7 - 4.7\,kpc.

We have also employed the empirical relation between the observed bolometric flux and the distance presented by \citet[][]{2005A&A...440..995S}. Using the relation for red giants \citep[eq. 27 in][]{2005A&A...440..995S} we then obtained the distance range of 6.9 - 7.8 kpc. Note that this relation is based on rather a small number of values. Therefore, its statistical significance is low, which can only give us a rough estimate.

The independent distance estimate can be obtained from the astrometric data published in the \textit{Gaia} DR2 and EDR3. There are some limitations due to the presence of the zero-point offset in parallaxes, which is a nontrivial function of the magnitude, colour and ecliptic latitude of the object \citep{2021A&A...649A...2L,2021ApJ...907L..33S}, due to the orbital motion of the long-period binaries \citep[][]{2019ApJ...874..178S}, and due to the fact that the reliable distance cannot be obtained by simple inversion of the parallax \citep{2018A&A...616A...9L}. The parallax of $(0.119 \pm 0.076)$\,mas and $(0.104 \pm 0.03)$\,mas was published for Hen~3-860 in the \textit{Gaia} DR2 and EDR3, respectively. The goodness-of-fit of 20.6 and 27.7 indicate a very poor fit to the data in both cases. \citet[][]{2018AJ....156...58B,2021AJ....161..147B} adopted the probabilistic approach with prior constructed from a 3D model of the Galaxy and obtained a distances of 5.1\,kpc (with uncertainty of 3.8 - 7.0\,kpc) and 6.4 kpc (5.7 - 7.4\,kpc) from the DR2 and EDR3 data, respectively. \citet{2019A&A...628A..94A} obtained the photo-astrometric distance of 5.4\,kpc (3.6 - 7.1\,kpc) using the StarHorse code and the data from \textit{Gaia} DR2 and photometric catalogues of Pan-STARRS1, 2MASS, and WISE.

All the employed methods resulted in more or less consistent results. In this research, we have adopted the distance of 5.0 kpc for the symbiotic system Hen 3-860.

\subsection{Hot component}

The hot components in burning symbiotic stars are typically too hot to be observable at optical wavelengths at which the giant and nebular emissions dominate \citep{2005A&A...440..995S}. To estimate the parameters of the hot component (the white dwarf with its pseudophotosphere) in Hen 3-860, we have used the indirect methods for assessment of the parameters of the central source of ionising photons based on the nebular emission line fluxes. Under the assumption of case B recombination and a black-body spectrum of the hot component, we can estimate the $T\rm_{h}$ and $L\rm_{h}$ using the fluxes of H$\beta$, \ion{He}{i} 5876\AA, and \ion{He}{ii} 4686\AA\,\,\citep{1981ASIC...69..517I,1997A&A...327..191M}. 

For the calculation, we have used the fluxes obtained from two low-resolution spectra calibrated to the ASAS-SN $g$ magnitude from the same time ($g$ = 14.6\,mag). Below, we list the resulting average values of parameters as the individual values obtained from these two spectra were very similar (the time difference between the two spectra is only 2\,days). 

The temperature $T\rm_{h} \sim 1.8 \times 10^{5}\,$K was obtained using the equation 2 of \citet{2016MNRAS.456.2558L}. The value estimated in this way is an upper limit and might be overestimated by 15-20\% \citep[see the discussion for AG Dra in][]{2017gacv.workE..60M}. The lower limit can be obtained from the maximum ionisation potential $IP$ derived from the spectrum and using the relation $T\rm_{h}$ [10$^3$\,K] $\sim$ $IP_{\text{max}}$\,[eV] \citep{1994A&A...282..586M}. In the case of Hen 3-860, the lines with the maximum $IP$ are that of [Fe VII] (the upper panels of Fig. \ref{fig:axper}), and the lower limit of the hot component temperature is $T\rm_{h} \sim 1 \times 10^{5}\,$K. We should note that these emission lines were detected only in our high-resolution spectra, and the spectra from 2020 were obtained with lower resolution which could prevent them being detected in the noise. Another possibility is that the emission lines [Fe VII] had not yet recovered after the outburst at the time of the 2020 observations, as  during the outbursts of symbiotic stars the highly ionised elements typically disappear from the spectra in response to the expansion and decrease of the hot component temperature.

The luminosities $L\rm_{h}(\ion{He}{ii}\,\lambda 4686, \ion{He}{i}\,\lambda 5876, \rm H\beta) \sim$ 1\,080\,L$_{\odot}$, $ L\rm_{h}(\ion{He}{ii}\,\lambda 4686) \sim$ 1\,260 L$_\odot$, and $L\rm_{h}(\rm H\beta) \sim 550\,$L$_\odot$ for the hot component of Hen 3-860 were calculated using equation 8 of \citet{1991AJ....101..637K} and equations 6~and 7~given in \citet{1997A&A...327..191M}, respectively. We should note, that the method of \citet[][]{1991AJ....101..637K} is a proxy only of the radiation emitted shortwards of 1200 \AA, but is well applicable in our case \citep[as was used for other symbiotic stars, see, e.g., ][]{2014MNRAS.444L..11M,2019AcA....69...25G} as the hot component, given its high temperature, radiates mostly in this part of the electromagnetic spectrum. The numbers of H$^0$ and He$^+$ ionising photons entering these equations, corresponding to $T\rm_{h} \sim 1.8 \times 10^{5}\,$K, were obtained from \citet[][]{1987A&A...182...51N}. Note that the low luminosity derived from the H$\beta$ flux is a result of significant central absorption of this emission line (visible in the high-resolution spectra in Fig. \ref{fig:axper}). Due to the fact that the part of the high-resolution spectrum covering H$\beta$ was rather noisy and the spectral line seems to be significantly underexposed, we have not used these data for analysis. At the same time, the central absorption of H$\beta$ is not detectable in the used spectra due to their low resolution and consequently, it was not possible to make a reliable correction for the central absorption. Moreover, it is not possible to estimate the correction for the low-resolution spectra from the high-resolution spectra as they were obtained at different times and in different phases of the recovery from the outburst.

Although according to \citet{1997A&A...327..191M}, these estimates have accuracies of a factor of $\sim$ 2, mostly due to uncertainties in reddening and distance, they allow us to compare the hot component of Hen 3-860 with the ones in other symbiotic systems (e.g., to distinguish between systems powered by thermonuclear burning or accretion only).Obtained temperatures ($> 10^5$\,K) and luminosities ($\sim 10^2 -  10^3$\,L$_\odot$) are typical for quiescent burning symbiotic stars (see Fig. 4 in \citealt{2003ASPC..303....9M} and also \citealt{1991A&A...248..458M,2007ApJ...660.1444S,2019arXiv190901389M}).

To summarise, the hot component of Hen 3-860 is a white dwarf with an extensive and dynamic pseudophotosphere. Based on the spectra obtained during the transition period, we have derived the parameters of the hot component: $T_{\rm h} \sim 1.8 \times 10^{5}\,$K and $L_{\rm h} \sim 1\,170\,$L$_\odot$. The derived radius of the hot component $R_{\rm h} \sim 0.035\,$R$_\odot$ (based on the black body assumption) is 2.5 times larger than the white dwarf radius for a mass of 0.5 M$_\odot$, a~typical mass of white dwarfs in symbiotic systems \citep[][]{2003ASPC..303....9M}, however, consistent with radii of other hot components in symbiotic stars \citep[e.g.,][]{1991A&A...248..458M,2017AJ....153..160S}. 

The (relatively cool) shell, which developed during the outburst and redistributed a fraction of the hot component radiation into the optical represents the eclipsed object in the symbiotic system \mbox{Hen 3-860}. The temperature of this shell is $T_{\rm s} \sim 5\,600 - 5\,900\,$K if we assume that it have same luminosity as the hot component and the radius $R_{\rm s} \sim 33 - 36\,$R$_\odot$ (see Sec. \ref{sec:orb}). Note that it is only a~rough estimate as the shell is actually only partially optically thick and the radiation of the nebula also contributed to the optical. The shell represents the central, densest part of the extended nebula ionised by the hot component which thickens and cools during outburst (significant eclipsing minima) and dissolves by subsequent expansion during the transitional period (gradual disappearance of eclipsing minima).

On the other hand, the shell that is optically thick enough to be subject to the observed eclipses would also obscure the ionising radiation of the hot component, the parameters of which were determined based on the emission lines formed in the ionised region surrounding this shell. If we assume the simultaneous presence of the hot ionising source and the relatively cool shell, this can indicate that the eclipsed object is not spherically symmetric. The most likely possibility is that it is a torus, thick enough in the orbital plane to be subject to the observed eclipses and at the same time open around its axis allowing the radiation of the hot component to escape and ionise surrounding nebula. Similar model was suggested for AX Per during its active stages \citep[][]{2011A&A...536A..27S}. The possibility that the optical properties of the eclipsed object are asymmetric and therefore vary with the orbital phase is not ruled out either, as the analysed photometric and spectroscopic observations were not obtained at the same time - the parameters of the eclipsed object and the hot component were obtained around phases 0.0 (JD\,2\, 458\,595 and  JD 2\,459\,197) and 0.76 (JD\,2\,459\,050), respectively. 

\subsection{Comparison with AX Persei} 
\label{sec:axper}
\begin{figure}
\begin{center}
\includegraphics[width=\linewidth]{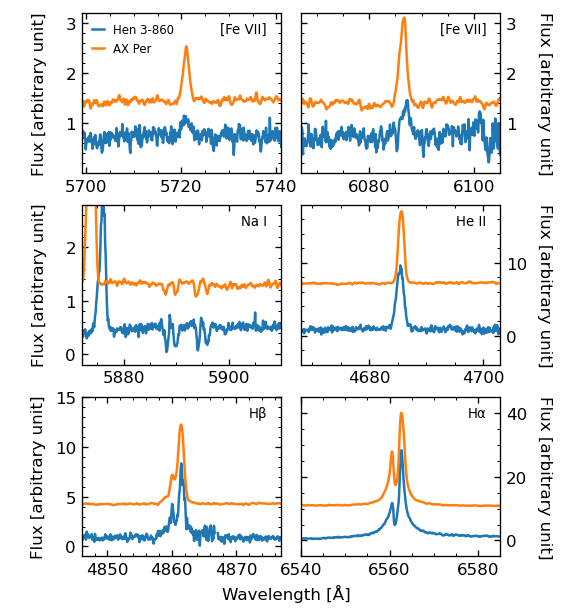}
\caption{Comparison of the profiles of selected spectral lines in our high-resolution spectrum of Hen 3-860 (blue; February 4, 2021) and in the spectrum of well-known symbiotic star AX Per (orange; November 5, 2020). The spectrum of AX Per was downloaded from the ARAS database \citep{2019CoSka..49..217T}.}
\label{fig:axper}
\end{center}
\end{figure}

As confirmed by the analysis presented in this article, Hen 3-860 belongs to the group of eclipsing symbiotic binaries. It is therefore tempting to compare it with another well-known eclipsing symbiotic star, AX Per. The components of AX Per are an M4.5 giant and a~white dwarf on 680-d orbit \citep[e.g.,][and references therein]{2011A&A...536A..27S}. Although the giant in AX Per is of a slightly later spectral type than that in Hen 3-860 (and therefore its radius is larger), both systems have very similar eclipsed objects - the radii obtained from the contact times are 33 - 36\,R$_\odot$ (Sec. \ref{sec:orb}) and 27 - 36\,R$_\odot$ \citep{2011A&A...536A..27S} in the case of Hen 3-860 and AX Per, respectively. 

Both stars manifest [\ion{Fe}{vii}] emission lines in their quiescent spectra, but do not show the Raman-scattered \ion{O}{vi} lines. They share a~similar H$\alpha$ line profile showing the blue-shifted absorption component, which probably originates in the wind of the cool giant (the lower right panel of Fig. \ref{fig:axper}). 

Furthermore, there is another noteworthy similarity between the spectra of Hen 3-860 and AX Per. Both symbiotic systems show two pairs of \ion{Na}{i} lines in their spectra (the lower left panel of Fig. \ref{fig:axper}). One pair has supposedly the interstellar origin and its velocity is used to put the lower limit on the distance to some symbiotic stars \citep[see e.g.,][]{2020A&A...644A..49M,2021MNRAS.504.2122M}. Our preliminary analysis of the spectroscopic data of AX Per confirmed that the other pair arises in the symbiotic system. During the active stages of AX Per, the emission doublet is sometimes observed instead of the absorption one. 

\begin{figure}
\begin{center}
\includegraphics[width=0.87\linewidth]{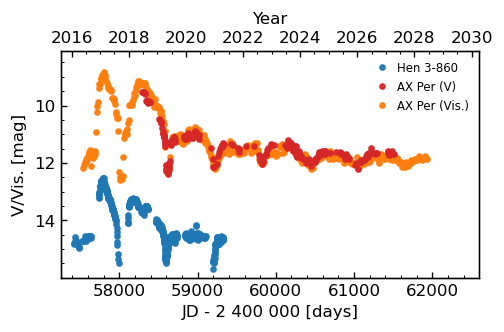}
\caption{Comparison of light curves of AX Per and Hen 3-860. The light curve of AX Per, covering the time interval of 1987 - 2001, is constructed on a basis of visual observations (smoothed within 10-day bins) from AAVSO database \citep[][]{kafka_2021} and $V$ data from \citet[][]{2001A&A...367..199S}. The light curve of AX Per was shifted in time and re-scaled to match the orbital period of Hen 3-860.}
\label{fig:axperphoto}
\end{center}
\end{figure}

In addition to above, there is striking similarity between the light curve of Hen~3-860 covering its recent outburst (2017 - 2019), and that of AX~Per during its outburst in 1987 - 1990. This may indicate a similarity of the geometry and location of the main source of the optical continuum in the post-outburst (transition) stage of both symbiotic stars. In Fig. \ref{fig:axperphoto}, we plotted both light curves together. Since AX~Per has a slightly longer orbital period (680\,d) than Hen 3-860 (602\,d), we re-scaled the light curve of AX~Per to match the orbital period of Hen~3-860. The light curve was than shifted in JD so that the eclipse times of both stars match. 

This comparison not only allows us to notice similarities in evolution during the outburst and shortly after it, but it also allows us to predict the evolution of Hen~3-860 in the coming years. On the AX~Per light curve, we can clearly see that during several orbital cycles after the outburst, the narrow eclipses (well observable in 1992, 1994) gradually changed into a sinusoidal variation typical for the quiescent periods of symbiotic binaries. Hen~3-860 is now in the transition period from outburst and if there is no further unexpected activity, we can assume that it will also continuously enter the quiescence and wave-like variability will be observable in its light curve.

\section{Conclusions}
In this research, we have analysed spectroscopic and multi-frequency photometric data of Hen 3-860, the object observed in the recent brightening and proposed to be a symbiotic star. Our analysis confirmed that Hen 3-860 is a classical symbiotic star of the S-type. The cool component is an M2-3 giant with $T_{\rm eff}$\,$\sim$\,3\,550\,K, $\log g$\,$\sim$\,1.0, radius 60 - 75\,R$_\odot$ and luminosity of 540 - 760\,L$_\odot$. The second component is a shell-burning white dwarf possessing a high temperature of $1-2 \times 10^5$\,K and luminosity of $\approx 10^3\,\rm L_\odot$. The recent light curve of Hen 3-860 confirmed that the object is a representative of a group of eclipsing symbiotic binaries. The presence of eclipses allowed to obtain the orbital period of the system of 602\,days. The symbiotic system experienced at least 4 outbursts in the last 120 years (1928, 1941, 2006, 2016-2019). Hen 3-860 is now in the transition period from the active stage to the quiescence. Based on its similarity to AX~Per, well-known eclipsing symbiotic binary we can assume that after few orbital cycles, the narrow eclipses in the light curve of Hen 3-860 will gradually change into wave-like variability typically observable in quiescent symbiotic stars. Therefore it is worth monitoring the system in order to document in detail the recovery of the system from the recent outburst during the transition into quiescence.  

\section*{Acknowledgements}

We are thankful to an anonymous referee for the comments and suggestions greatly improving the manuscript. We acknowledge and thank T. Lester (ARAS Group) whose spectrum of AX Per is used in this paper. We are grateful to F. Teyssier for coordinating the ARAS Eruptive Stars Section. This research was supported by the \textit{Charles University}, project GA UK No. 890120, the internal grant VVGS-PF-2021-1746 of the \textit{Faculty of Science, P. J. \v{S}af\'{a}rik University in Ko\v{s}ice}, and the \textit{Slovak Research and Development Agency} under contract No. APVV-20-0148. The research of MW is supported by the grant GA19-01995S of the Czech Science Foundation. The DASCH project at Harvard is grateful for partial support from NSF grants AST-0407380, AST-0909073, and AST-1313370. 

\section*{Data Availability}

Low-resolution spectra used in the paper are available in the ARAS Database. Photometric data are accessible from the websites of the surveys and from the DASCH archive. The other data are available on reasonable request to the authors.



\bibliographystyle{mnras}
\bibliography{example} 


\vspace{10cm}
\appendix

\begin{table}\section{Log of observations}
\label{obs_log}
\caption{Log of spectroscopic observations. Observatory codes: iTSS - the iTelescope facility at Siding Springs, Australia; MIA - the private station at Mirranook Armidale, Australia; CTIAO - the Cerro Tololo Inter-American Observatory, Chile.}             
\label{table:log_obs}      
\centering
\begin{tabular}{lllcc}
\hline\hline
JD  2\,459.. & Date    & Res. & $\lambda_{\rm min}$-$\lambda_{\rm max}$ [\AA] & Obs. \\
\hline
049.4     & July 18, 2020 & 1176      & 3800-7501                                 & iTSS  \\
051.5     & July 21, 2020 & 1778      & 4000-8000                                 & MIA  \\
249.9     & February 4, 2021 & 25\,000   & 4500-8500                                 & CTIAO  \\
294.6     & March 21, 2021 & 25\,000      & 4500-8500                                & CTIAO  \\
\hline
\end{tabular}
\end{table}


\bsp	
\label{lastpage}
\end{document}